# Introduction of 6$^{th}$ Generation Smart Phone combining the features of both Apple and Android smart phones.

[1]Nyembo Salama,[2]Christian Bach
Computer Science and Engineering Department
University of Bridgeport, Bridgeport, Connecticut 06604, USA
[1]**nsalama &[2]cbach{@bridgeport.edu}**

**Abstract:**In this paper, we present our novel contribution methodology based on the results of case study that has been implemented in our research environment to test with new technique the usability of both intelligent Android and Apple phones. This analysis of the case study stands for features similar to applications, operating system, hardware and software structure, battery life, and online based websites. Multiple interrogations were applied to collect user answers ongoing features. Users directly react by responding based on their daily used product experience. Consequently, the estimation is based on the data that has been unregistered from the user. The most recent results will end up by introducing a combination of ideal features on both products to build a wonderful extended product in the future.

**Keywords:** Smart Phone, Android based smart phone, Apple based smart phone, Usability testing.

## 1. INTRODUCTION

Early exploit of mobile devices implicated the model of using phones to secure communications over the network. The innovation technique and technology has authorized us to obtain the high stage level that mobile devices presently hold the capability to sprint desktop applications. For mobile devices to fill up the condition of this development, mobile hardware automatically has to be adapted to handle these modifications such as long battery life [7]. Smart phones enjoy implementation of serial features such as multitasking and touch screen. But, the procedure of the implementation completely remain different from one mobile device product to another which is characterized by the way that is planned to please the user needs [8]. Worldwide, people are actually interested, and active users of smart phones, and are also enjoyed the benefits of applying applications created by software developers. The creation of best products, design reengineered style, online stores capabilities, and applications build the highest degree of competition between smart phones companies. Android has more applications. In addition, Apple has focused on the design reengineered style and online stories.

Owing to the continuously growing nature of smart phone, the difference between smart and dumb phones is uncertain. In reality, even dumb phones possess some features of smart phones, such as touch screen and operating system. At the conclusion of research [11 defines that smart phone as works an open operating system and is enduringly associated to the Internet [11].

Most people are now active users of smart phones and are availing the benefits of using applications built by software developers.

The contest between smart phones, the several companies have introduced several products especially, design style, online stores, and applications [12]. Therefore, Apple has paid attention on design style and online stores but, Android has developed more applications.

## 2. RESEARCH METHODOLOGY

One of the challenging tasks is to compare between two leading products. On the base of comparison, one better product with ample features is selected but no one product can meet the requirements of end users. The needs of users are variable with passage of time to fulfill their functional requirements. To meet needs of users, we apply usability testing method, complete survey about both Android and Apple iphone.

The testing method and survey are also supported with heuristic loom to design new model that helps to resolve sufficient requirements of users in real environment. However, the appropriateness of this model needs more focus and contemplation.

### 2.1.Setting goals

We Study almost all features of Android and Apple iphone that help to build new model for mobile phone. On the basis of survey, we conduct the usability testing. We also analyze main features of both leading mobiles and arrange these features into five categories to build better model. The features of mobile phones consist of operating system, integrated technology, running applications, multitasking /multimode support and compatible framework. These five categories help to understand and erect new model. Our new developed model is highly compatible with the needs of end users.

### 2.2.Data Gathering

The collection of data follows the standards of fieldwork research explained in [10]. Several data sources are applied that covers interviews, observations and literature. The authors have also research capability in industry to trigger the data to navigate the interview process. As, the first stage, we collect the relevant features of Android and Apple iphone. In second phase, we sort out the best features of each mobile phone. Finally, we get feedback through observation, interviews and coordination with skillful persons of related filed in order to obtain helpful suggestions.

### 2.3.Interviewing process

Interviews are performed with administrative staff, experts, student-researchers, teaching staff and industrialist [9]. The time duration for each interview consists of approximately 20 minutes and performed on a one-to-one basis.

### 2.4.Power of Case study

The use of data collections, observations, interviews and documentary sources are purely based on scientific reliability. The validity of each foundation must be highlighted. Each interview is recorded and then transcribed. The most noteworthy discussions are taken to make sure that collected data is converged on corresponding facts, as described [5].

### 2.5.Evaluation of study

It is done on basis of testing methods. We finish the process of testing then make observation. In our case, we give the questionnaire to students for obtaining feedback given in table 1. Each student gives feedback for each feature of android and apple smart phone.

Table 1: Showing feedback obtained from 44 students

| **Name of features** | **Name of Mobiles** | **Strongly Agree** | **Agree** | **Neutral** | **Disagree** | **Strongly Disagree** |
|---|---|---|---|---|---|---|
| 1.Is operating system supporting several types of developed applications? | Android | 19 | 18 | 06 | 01 | 00 |
| | Apple Iphone | 08 | 12 | 20 | 04 | 00 |
| 2.Is dual core A5 chip available? | Android | 17 | 15 | 12 | 00 | 00 |
| | Apple Iphone | 23 | 12 | 09 | 00 | 00 |
| 3.Can we get faster page loading and improved graphical performance | Android | 16 | 20 | 08 | 00 | 00 |
| | Apple Iphone | 26 | 14 | 04 | 00 | 00 |
| 4.Is this providing faster page transfer from Iphone to computer? | Android | 18 | 20 | 06 | 00 | 00 |
| | Apple Iphone | 20 | 14 | 10 | 00 | 00 |
| 5.Is this multitasking? | Android | 36 | 04 | 04 | 00 | 00 |
| | Apple Iphone | 10 | 32 | 02 | 00 | 00 |
| 6.Is framework compatible with all existing apps of Google and Microsoft? | Android | 30 | 10 | 03 | 01 | 00 |
| | Apple Iphone | 04 | 20 | 04 | 16 | 00 |
| 7. Is it using freely any number of third party applications? | Android | 32 | 10 | 02 | 00 | 00 |
| | Apple Iphone | 12 | 19 | 13 | 00 | 00 |
| 8. Is it giving multi application support? For instance, if you listening songs and somebody calls you then song will be paused until call finishes | Android | 12 | 30 | 02 | 00 | 00 |
| | Apple Iphone | 32 | 09 | 03 | 00 | 00 |
| 9.Is it adjusting display automatically? | Android | 14 | 15 | 10 | 05 | 00 |
| | Apple Iphone | 20 | 19 | 03 | 02 | 00 |
| 10.Is it providing real time chatting? | Android | 39 | 05 | 00 | 00 | 00 |
| | Apple Iphone | 39 | 04 | 01 | 00 | 00 |
| 11.Is it responsive touch screen? | Android | 22 | 16 | 06 | 00 | 00 |
| | Apple Iphone | 27 | 10 | 07 | 00 | 00 |
| 12.Is it easy to set up process? | Android | 12 | 10 | 22 | 00 | 00 |
| | Apple Iphone | 12 | 12 | 20 | 00 | 00 |
| 13.Is it providing attractive game features? | Android | 38 | 02 | 04 | 00 | 00 |
| | Apple Iphone | 40 | 02 | 02 | 00 | 00 |
| 14.Is it providing Email, browsing, VOIP, SSH, VPN support? | Android | 34 | 10 | 00 | 00 | 00 |
| | Apple Iphone | 32 | 12 | 00 | 00 | 00 |
| 15.Does it have accelerometer support? | Android | 23 | 17 | 04 | 00 | 00 |
| | Apple Iphone | 31 | 09 | 04 | 00 | 00 |
| 16.Does it have GPRS support? | Android | 44 | 00 | 00 | 00 | 00 |
| | Apple Iphone | 44 | 00 | 00 | 00 | 00 |
| 17.Does it have battery life tracker supported application? | Android | 14 | 10 | 20 | 00 | 00 |
| | Apple Iphone | 22 | 10 | 12 | 00 | 00 |
| 18.Is it using own operating system? | Android | 44 | 00 | 00 | 00 | 00 |
| | Apple Iphone | 00 | 00 | 40 | 04 | 00 |
| 19.Is it using excellent inter-process architecture? | Android | 32 | 10 | 02 | 00 | 00 |
| | Apple Iphone | 12 | 19 | 13 | 00 | 00 |
| 20.Does it have intelligent assistant support? | Android | 12 | 10 | 22 | 00 | 00 |
| | Apple Iphone | 22 | 17 | 05 | 00 | 00 |
| 21.Does it provide creation of own distribution channels? | Android | 12 | 30 | 02 | 00 | 00 |
| | Apple Iphone | 12 | 12 | 20 | 00 | 00 |
| 22.Is it providing muti-application support? | Android | 22 | 20 | 02 | 00 | 00 |
| | Apple Iphone | 23 | 17 | 04 | 00 | 00 |
| 23.Does it have 4G Capable with Speeds up to 42Mbps? | Android | 40 | 04 | 00 | 00 | 00 |
| | Apple Iphone | 40 | 03 | 01 | 00 | 00 |
| 24.Does it have 8 Megapixel Camera? | Android | 44 | 00 | 00 | 00 | 00 |
| | Apple Iphone | 44 | 00 | 00 | 00 | 00 |
| 25.Is it Dual core product with 1.5 GHz Processor? | Android | 32 | 07 | 05 | 00 | 00 |
| | Apple Iphone | 28 | 10 | 06 | 00 | 00 |

We hereby combine the best features of Apple and Android smart phones and recommend design of new powerful smart phone given in figure 1.

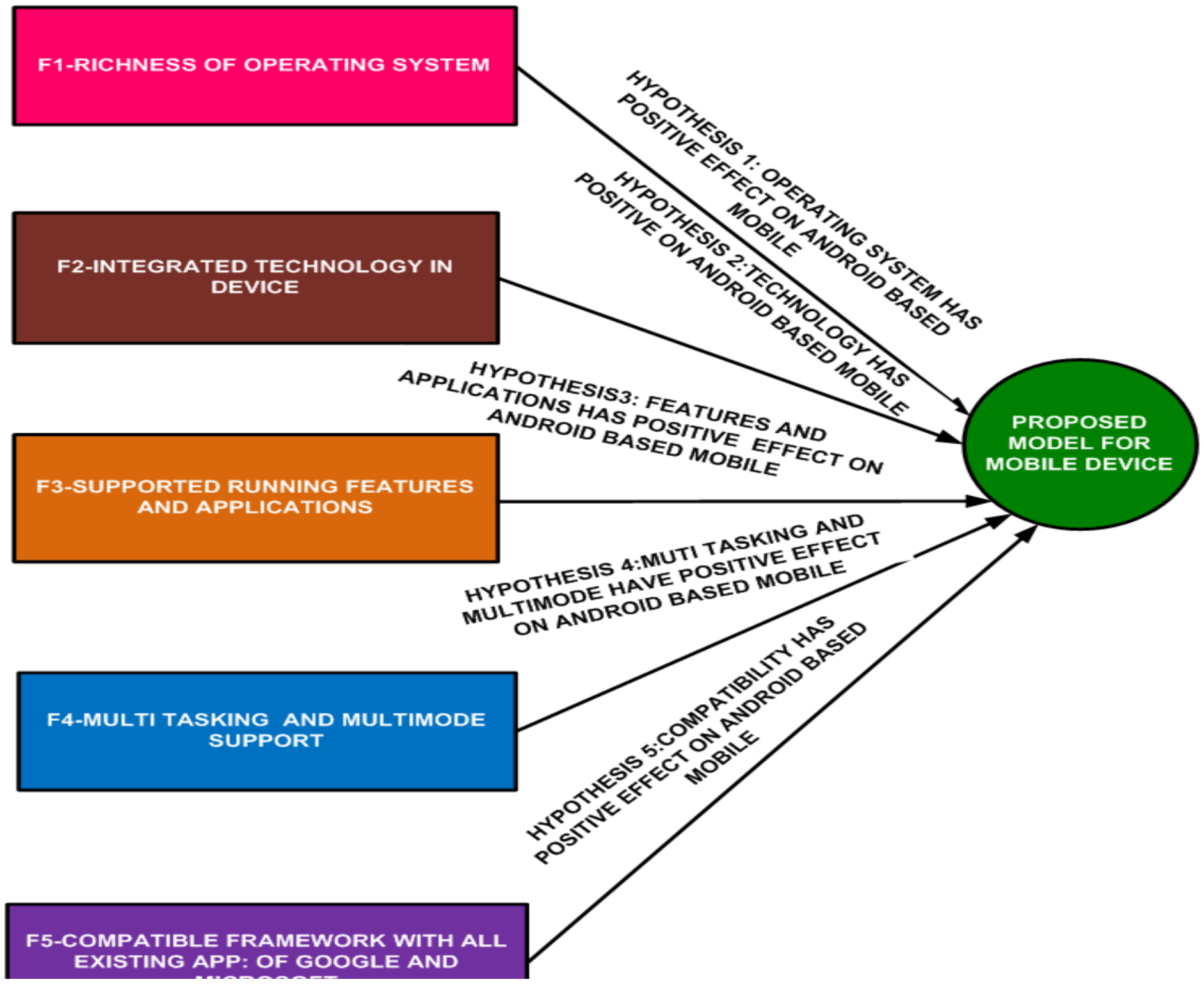

Figure 1: New proposed model for mobile device

## 3. DISCUSSION AND DESIGN OF NEW MODEL

We hereby discuss new proposed model that integrates the features of android and apple smart phone.

### 3.1.Integrated technology in device

The integrated device technology concerns about the system architecture. In this design, the developed system integrates heterogeneous network management technologies, which also include the Android mobile system software design. A peculiar feature of the system developed is that it can be useful as Home Gateway Entertainment (HGE), Vehicle On-Board Unit (OBU), and portable network device unit. The system provides convenient wired and wireless internets, which can be effectively accessed by users.

A typical application of Android is the HID. The system provides two functions; which are roaming and sharing, and the interface has been equipped with heterogeneous network interface. Roaming has the capability of searching and choosing best resources wireless network-an example is RSS, and Bandwidth which are used by users, based on the pertaining conditions. When roaming is desired in the heterogeneous and homogenous network, this becomes most useful. Based on the Sharing functionality of the system, the system is capable of allocating resources to all users, with the intention of the system having been interconnected with wired/wireless multiple network.

People and users alike normally encounter sharing function, especially when they are home or in tall buildings, and secluded place. Also, with users and people who are outside home and buildings, roaming functionalities uses intelligent Mobile Internet Device (MID). User traffic demands can be met when the system has chosen more or available network resources [1].

### 3.2.Supported running features and application

The android mobile device system development also supports contents modules used for description of the supported feature, which runs on the Android system application. An application software known as the Java Native Interface (JNI) is a programming software framework, which runs on the Java Virtual Machine (JVM), which normally display two functions: to call and to be called by native application, written in C/C++ assembly language. The Android machine also requires the JNI interface for an application. When necessary, and it is required Android application developers should resort to use of Android NDK, embedded in code native libraries as the required application development platform; then the application which is running on the virtual machine in the JNI interface can be used to call the native function. An important large part in native code, written in the language C/C++ construction, which is also used to access the devices, platform specific task, which also enhances the application by use of critical code, can be developed by the JNI reuse interface [2].

### 3.3.Multi-tasking and multimode Support

We refer here multitasking and multimode support to portable Lab system development of the Android mobile device system application. This portable Lab system development has a main requirement which needs to be followed in our system development, which supports the architectures and the technologies used in our system implementation. The established principles and requirements about implementing our new methodology and design for a new system will include but not limited to the following [6].

i. User account creation and its validation for academicians which include professors and students
ii. Mobile device system application connection to remote server
iii. Data extraction from server by the data acquisition board, which must include a database to store data collected.
iv. There must be a database server linked to our system application of the mobile device.
v. Information reproduction system of our mobile device from database data

Choice of Online/Offline mode made possible and must be operated from the remote database.

vi. Local and remote database must be synchronized together.
vii. Information visualized must be at the harmonic power and voltage values level which must include graphics.

viii. The server side must be able to search power and read which uses different criteria like date and time.
ix. It must be able to annotate/comment/ and be questioned on data readings, which can also be made to view specific data comments.
x. It must be deployed in Android platform.

### 3.4.Richness of operating system

The development of the Android software operating system can be implemented on a hardware platform, and also, telemetric scenario can be included. The detail of the major features of the operating system is presented here. Consideration of Linux is essential for the built of the Android features source files. Linux or Mac operating system (Mac OS) is included. Also, in consideration of that, Linux Fedora and gcc version 4.3.2 is the building environment for the Android device.

Not distending, we required other software packages such as: Java JDK, Python, and Repo to be all inclusive. The Android mobile device system is configured to theX86 architecture, since the OS required that. Also, another pertinent factor is to build the kernel of the Android device based on the Linux version 2.6 and the latest. This is required to provide system services such as memory management, security, network stacks, process management and other driver models.

It is essential to drive the peripheral of the hardware device, which normally makes it possible to run Android OS on the target device. The Android kernel source already includes drivers, and chooses functions that also comply with that; the regular thing done is that, the drivers are added in the kernel. Some dependent driver functions such as touch screen and power management, 3.50 modules USB serials, protocol stacks and network functions add and modify the right driver source, which validly drives the target device. Software driver collocating tools that work best on 802.11 Wireless LAN cards require configuration tools. Also, the 3.50 USB modem requires 3.50 dialer, and OPS navigator tools also require library relations.

The kernel files are compiled to form the kernel image after its source is configured. Android mobile device system fundamental library depends on its operating core, which is fully capable of some functionality as; its surface management, media support, system library, database engine. Also, certain functions such as Handover, QoS and OPS are normally implemented at libraries defined by each user. This has the capability of enhancing network connections and services, which also has the attribute of better network quality resources based on its user services.

Android system and its network functions require external tools in its configuration; and the functions are also linked with the kernel and its application layer. This helps adding program codes to the Android system files. Some of these system file are wireless configuration tool files, network tool files dialer tool files, and Busy box command tool files. Application framework functions source code some of which are Resource Manager and Windows Manager and many others, also need to be modified [3].

Typical function such as the curs or is integrated to the application framework function in the system designed. The intention is to make controlling of the Android system more suitable for remote controlling by users.

In this mobile system application development based, Android SDK (i.e. Software Development Kit) is the useful tool. The internet serves as a means by which most services will obtain its data information, using wireless devices. Based upon all these preceding described factors, network handover management services are considered more important, which also enhances communication service qualities. Relative libraries are necessary that serves as a reference in Android application development system files. The android OS has been installed successful to the target device, as a result of the application developed and also downloaded as APK file [4].

### 3.5.Framework with existing applications of Google and Microsoft

There must be compatibility in all existing framework which Android supports with all existing application types (Microsoft and Google etc) .Originally, Android is required to provide multimedia framework software support stack. This vision of Android is necessary in order to integrate all software modules, device drivers and their algorithms. This will support multimedia application convergence standard which can easily transplant. Since high performance is desired, and for the reason of low power embedded multimedia framework, large computational load of playback could be realized. In order to be able to compute multimedia data digital signal processor (DSP) will be required.

Moreover, codec which encode/decode video and audio multimedia is also an urgent requirement for this design. Multimedia framework is therefore requires support from encode/decode, or in other words from the codec [2].

## 4. FINDINGS

The findings are based on testing procedure that invites the participants belonging to different background. Some are familiar and belonging to mobile and wireless communication field and fewer possess less expertise in this field but know how to use mobile devices. The group leader who is much familiar with this field records the performance for the features of Interactive architecture and applications on the basis of activities performed during the testing.

The performance analysis requirement or beta testing method also involves three steps. First, we introduce the testing procedure from design phase to conducting the test. Second, make all the related operations of architecture and applications and finally, we give the questionnaire to all the participants based on 5-level Likert method. On the basis of feedback, we collect the following statistical data that will be applied for developing innovative mobile applications to support pedagogical activities given table 1.

## 5. CONCLUSION

The usability testing of a consumer product has more significance for introducing new technology that is better than existing technology. At the completion of usability testing, we are capable to determine the differences between Android and Apple mobiles. In addition, we are able to produce an ideal model for sixth generation of mobile phone. Results point out that performing of time-consuming usability test will improve the flaws of product and increase user interaction.

On basis of result, we propose the most dominant process to introduce ideal smart mobile phone with many extra features.